\documentclass[aps,pra, twocolumn]{revtex4}

\usepackage[normalem]{ulem}
\usepackage{newlfont}
\usepackage{color}         
\usepackage{amssymb}
\usepackage{amsfonts, cleveref, varioref, nameref,soul}
\usepackage{amsmath}
\usepackage{wasysym}
\usepackage{graphicx,subfig}
\usepackage{epsfig}
\usepackage{amsthm}
\usepackage{bm}
\usepackage{mathrsfs}
\usepackage{epstopdf}
\usepackage{theorem}
\makeatletter
\newcommand{\Rmnum}[1]{\expandafter\@slowromancap\romannumeral #1@}
\makeatother
\usepackage[colorlinks=true, citecolor=blue, urlcolor=blue ]{hyperref}

\begin{document}

\title{Quantum Process Randomness}



\author{Sreetama Das\(^1\), Asutosh Kumar\(^{2}\), Aditi Sen(De)\(^1\), Ujjwal Sen\(^1\)}

\affiliation{\(^1\)Harish-Chandra Research Institute, HBNI, Chhatnag Road, Jhunsi, Allahabad 211 019, India}
\affiliation{\(^2\)P.G. Department of Physics, Gaya College, Magadh University, Rampur, Gaya 823 001, India}

\begin{abstract}

We introduce the concept of ``randomness'' of a quantum process by examining a 
measurement to estimate the figure of merit of the process.
We quantify the randomness by the standard deviation 
of the probability distribution obtained from the measurement,
and 
inquire whether an optimization of the quantum process based on that measure, can reach the point where the process operates with maximum fidelity. We consider approximate quantum cloning and teleportation processes to illustrate the concept.
Interestingly, we find that the optimal approximate state-dependent quantum cloning machine obtained by maximizing the fidelity is different from that obtained by minimizing the randomness.

\end{abstract}

\maketitle


%
%
%
%

The distinctive features of quantum mechanics, uncustomary from a classical perspective, have led to path-breaking successes in several spheres, including in quantum information. It has been possible to propose quantum processes and build quantum machines corresponding to several of them, which can perform tasks with enhanced efficiencies, as compared to their classical counterparts. Examples of such quantum processes are quantum key distribution \cite{ekert}, quantum dense coding \cite{gamanush}, quantum teleportation \cite{bennett},  
 Shor algorithm \cite{jati}, etc. 
Any quantum process is inevitably subject to environmental noise, which prohibits it from being executed perfectly. It causes the quantum process to produce outputs which differ from the desired ones. Also, in some scenarios, it may in principle be impossible to achieve a particular task through a quantum process, reasons for which being inherent in the quantum mechanical laws. A prominent example is the impossibility of exact cloning of an arbitrary  quantum state as well as that of a non-orthogonal pair of states \cite{brisTi-poRe-Tapur-Tup}. In these scenarios, we try to optimize the system parameters in the best possible way so that the state obtained at the output of a quantum process remains as close as possible to the actual output targeted. To measure how close the obtained output state is to the desired state, one typically uses a fidelity measure between the two states. We argue that along with the fidelity, which can be seen as an average in a quantum measurement, the spread of the outcomes in the same measurement is also a physically relevant quantity. 

In this paper, we provide a formal definition of the spread, terming it as ``quantum process randomness", 
by using the statistical measure of standard deviation. 
We determine this quantity in universal and state-dependent approximate quantum cloning and approximate quantum teleportation. In particular, we find that the best quantum cloning machine in the state-dependent case when we maximize the fidelity is different from the one obtained by minimizing the process randomness. Minimizing the ratio of fidelity to the process randomness leads to the same machine that minimizes the process randomness.

Consider a quantum system for which the associated complex Hilbert space is \({\cal H}\). 
We define a ``protocol'' $\mathtt{P}$ as a rule that assigns a pure quantum state $\left|\varphi\right\rangle \in {\cal H}$ to another pure quantum state 
 $\mathtt{P}(\left|\varphi\right\rangle \left\langle \varphi \right|) \equiv 
\left|\tilde{\varphi}\right\rangle \left\langle \tilde{\varphi} \right|
$. Note that the output belongs to a Hilbert space, \(\tilde{\mathcal{H}}\), that can be different from that of the input.
The considerations can be generalized to  allow mixed quantum states as inputs and outputs in the protocol. 
It is to be remembered that \(|\tilde{\varphi}\rangle\) is a function of \(|\varphi\rangle\).
Note however that the protocol $\mathtt{P}$ may not be quantum mechanically allowed, at least not for all pairs of \((|\varphi\rangle, |\tilde{\varphi}\rangle)\). 
Let $\Phi$ be an allowed quantum mechanical ``process'' such that
\begin{equation}
\label{eq:process-action}
\Phi : 
\left|\varphi\right\rangle \left\langle \varphi \right| \longrightarrow
\rho_{\varphi}.
\end{equation} 
Again, the output state, \(\rho_\varphi\), of the process, \(\Phi\), is defined on the Hilbert space, \(\tilde{\mathcal{H}}\), which can be different from \(\mathcal{H}\), 
and again it is possible to generalize to the case of a mixed quantum state input to the process.
The actual output, \(\rho_\varphi\), can be different from the desired one, 
\(|\tilde{\varphi}\rangle \langle\tilde{\varphi}|\) of the protocol \(\mathtt{P}\).

The quantum process \(\Phi\) is associated with the protocol \(\mathtt{P}\) in the sense that the former approximates the latter as far as is quantum mechanically 
allowed. As an example, we can consider the \emph{protocol} of exact cloning of arbitrary quantum states of a certain \(d\)-dimensional Hilbert space -- known to be 
quantum mechanically impossible \cite{brisTi-poRe-Tapur-Tup} -- and the associated quantum \emph{process} of approximate universal quantum cloning \cite{jole-bhijte-moja-khub, bruss}. 
As a further example, we can refer to the \emph{protocol} of exact teleportation \cite{bennett} of arbitrary quantum states of \(\mathbb{C}^d\) by using a 
state that is not maximally entangled in \(\mathbb{C}^d \otimes \mathbb{C}^d\) and a finite amount of classical communication -- known to be 
disallowed in quantum mechanics -- and the associated quantum \emph{process} of approximate universal quantum teleportation through non-maximally entangled states and 
an additional usage of a finite amount of classical communication \cite{shatadru}. We now proceed to quantify the extent to which the quantum process \(\Phi\) can 
approximate the protocol \(\mathtt{P}\).

The usual quantification of the success of a process \(\Phi\) in simulating the protocol \(\mathtt{P}\) for the input \(|\varphi\rangle\) is via the 
``fidelity'',
\begin{equation}
\label{eq:fidelity}
F_{\Phi}(\varphi) = \langle \tilde{\varphi}| \rho_{\varphi}| \tilde{\varphi} \rangle.
\end{equation}
As $F_{\Phi}(\varphi)$ is a measure of how close the obtained state is to the desired one, we call it the ``efficiency'' or strength of the
process $\Phi$ for the state $\varphi$. Clearly, $0 \leq F_{\Phi}(\varphi) \leq 1$.  
This quantity provides an important piece of information about the success of the process in attaining the task it has been pressed into, viz. producing a state that is as close as 
possible to \(|\tilde{\varphi}\rangle\). A measurement of the quantity can be performed in the following way, which also gives a possible 
interpretation for \(F_{\Phi}(\varphi)\). 
Suppose that \(\mathcal{M}(\rho_{\varphi})\) forms 
the eigenbasis of \(\rho_\varphi\). If \(\rho_\varphi\) has full rank, \(\mathcal{M}(\rho_{\varphi})\)  spans \(\tilde{\mathcal{H}}\). Otherwise, 
\(\mathcal{M}(\rho_{\varphi})\) can be completed to get a complete orthonormal basis 
\(\mathcal{M}^c(\rho_{\varphi}) = \{|m^c_i(\rho_\varphi)\rangle\}_{i=1}^{\dim {\tilde{\mathcal{H}}}}\) of \(\tilde{\mathcal{H}}\). 
Consider the projection-valued measurement on such a 
complete orthonormal basis of \(\tilde{\mathcal{H}}\), that is partly or entirely the eigenbasis of \(\rho_\varphi\), and suppose that the state \(|\tilde{\varphi}\rangle\)
is fed as an input to that measurement. Let \(\lambda_i(\rho_\varphi)\) be the eigenvalue of \(\rho_\varphi\) corresponding to the 
eigenvector \(|m^c_i(\rho_\varphi)\rangle\). Of course, \(\lambda_i(\rho_\varphi) =0\) for \(|m^c_i(\rho_\varphi)\rangle \in 
\mathcal{M}^c(\rho_{\varphi}) \backslash \mathcal{M}(\rho_{\varphi})\). The average \(\sum_{i=1}^{\dim{\tilde{\mathcal{H}}}} \lambda_i(\rho_\varphi) p_i \), with 
\(p_i = |\langle \tilde{\varphi} | m^c_i(\rho_\varphi)\rangle |^2\) being the Born probability that \(|m^c_i(\rho_\varphi)\rangle\) clicks in the measurement, 
is 
the quantity of our interest, \(F_{\Phi}(\varphi)\). 
In the special case when $|\tilde{\varphi}\rangle = |\varphi\rangle$ (this is the case in the quantum teleportation ``protocol'' alluded to above), 
\(F_{\Phi}(\varphi) = \langle \varphi| \rho_{\varphi}| \varphi \rangle\).

Remaining still with the general case when $|\tilde{\varphi}\rangle = |\varphi\rangle$ is not necessarily true, although the 
quantity
\(\sum_{i=1}^{\dim{\tilde{\mathcal{H}}}} \lambda_i(\rho_\varphi) p_i \) 
contains 
important 
information about the experiment, 
there 
may still remain
further information 
to be gathered from the distribution \(\{p_i\}\).
In particular,
the spread of the distribution, which can, for example, be quantified by the standard deviation, \(Q_{\Phi}(\varphi)\), given by
\begin{equation}
\label{kolur-bolod}
 Q_{\Phi}^2(\varphi) = 
\sum_{i=1}^{\dim{\tilde{\mathcal{H}}}} \lambda^2_i(\rho_\varphi) p_i - \left(\sum_{i=1}^{\dim{\tilde{\mathcal{H}}}} \lambda_i(\rho_\varphi) p_i\right)^2, 
\end{equation}
contains a significant fragment of information about the distribution.
It 
can be rewritten as 
\begin{eqnarray}
\label{dhush-sssala}
Q_{\Phi}^2(\varphi) = 
\langle \tilde{\varphi}| \rho^2_{\varphi}| \tilde{\varphi} \rangle - \langle \tilde{\varphi}| \rho_{\varphi}| \tilde{\varphi} \rangle^{2} \nonumber \\
=\langle \tilde{\varphi}| \rho^2_{\varphi}| \tilde{\varphi} \rangle - F^2_{\Phi}(\varphi),
\end{eqnarray}
with \(F_{\Phi}(\varphi)\) being given in Eq. (\ref{eq:fidelity}). There are of course higher moments of the probability distribution that could have also been used, to obtain complete information about the 
process. However, in this paper, we restrict 
ourselves to the second moment.
Note that \(Q_{\Phi}(\varphi)\) quantifies a randomness that is inherent in the quantum process even for a given input. It is therefore a 
randomness of the process \(\Phi\) for a given input \(|\varphi\rangle\). We refer to it as the \emph{quantum process randomness} of the process \(\Phi\)
for a given input \(|\varphi\rangle\). 

\begin{figure}[!tbp]
\centering
\includegraphics[width=0.2\textwidth]{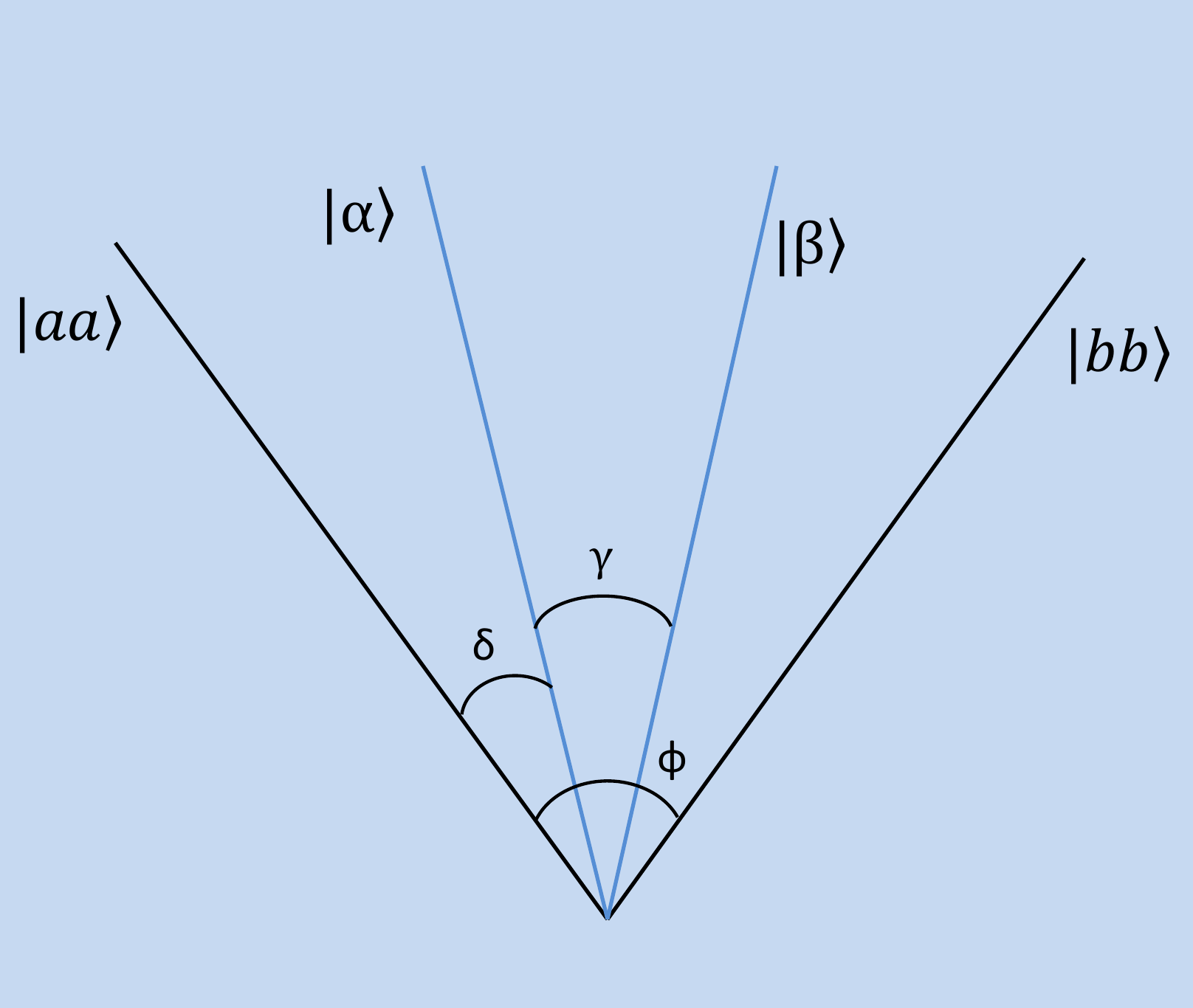}\label{fig:f1}
  \hspace{0.5cm}
 \includegraphics[width=0.2\textwidth]{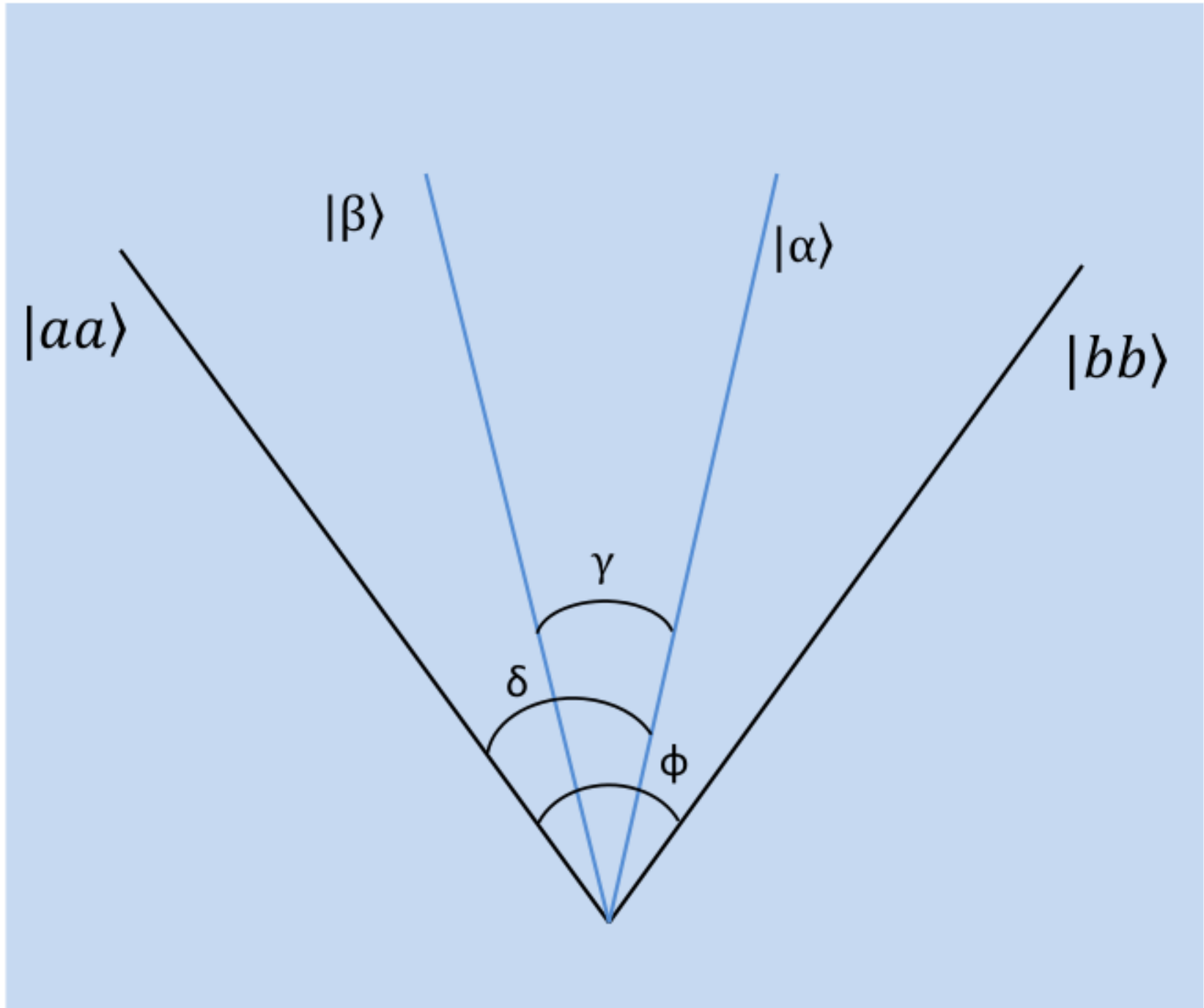}\label{fig:f2}
\caption{State-dependent quantum cloning. The states $ |aa\rangle $ and $|bb\rangle$ before cloning and the states $ |\alpha\rangle $, $|\beta\rangle$ after cloning remains in one plane, and are pictorially represented in the diagram. See Ref. \cite{bruss}. In the figure on the right, the positions of 
\(|\alpha\rangle\) and \(|\beta\rangle\) have been interchanged with respect to the one on the left.}
\end{figure}

In the special case when $|\tilde{\varphi}\rangle = |\varphi\rangle$,
we get
\begin{equation}
\label{eq:randomness}
Q_{\Phi}(\varphi) = \sqrt{\langle \varphi| \rho^2_{\varphi}| \varphi \rangle - \langle \varphi| \rho_{\varphi}| \varphi \rangle^{2}}.
\end{equation} 
If further, $\rho_{\varphi}\) is a pure state \(|\chi \rangle \langle \chi |$, the standard deviation 
reduces to
\begin{align}
Q_{\Phi}(\varphi) &= \sqrt{| \langle \varphi | \chi \rangle |^2 (1 - | \langle \varphi | \chi \rangle |^2)}.
\end{align}
Thus, $Q_{\Phi}(\varphi)$ vanishes when $|\chi\rangle$ is either orthogonal or equal (up to a phase) to $|\varphi\rangle$, 
and is nonvanishing otherwise. The cases when the output is orthogonal or equal to the input, there is of course no randomness, as only one of the 
``pointers'' clicks for the measurement in the basis 
\(\mathcal{M}^c(\rho_{\varphi})\).


In general, we may not be in a position to know the input exactly. Let \(\mathcal{S}\) (\(\subseteq \mathcal{H}\)) be the set that encapsulates the information 
about the input state \(|\varphi\rangle\), i.e., \(|\varphi\rangle \in \mathcal{S}\). In such a situation, we will have to consider the 
\emph{average quantum process efficiency}, 
\begin{equation}
\label{tora-chhai-kha}
\overline{F}_{\Phi} = \int_{\varphi \in \mathcal{S}} \langle \tilde{\varphi}| \rho_{\varphi}| \tilde{\varphi} \rangle d \varphi,
\end{equation}
and the \emph{average quantum process randomness}, 
\begin{equation}
 \label{ke-dil-abhi-bhara-nehi}
\overline{Q}_{\Phi} = \int_{\varphi \in \mathcal{S}} Q_{\Phi}(\varphi) d \varphi,
\end{equation}
where \(d\varphi\) is a Haar uniform measure on the set \(\mathcal{S}\), and where \(Q_{\Phi}(\varphi)\) is given in Eq. (\ref{kolur-bolod}).
For a given protocol, \(\mathtt{P}\), the main objective is to 
maximize \(\overline{F}_{\Phi}(\varphi)\) and to minimize \(\overline{Q}_{\Phi}(\varphi)\).

\emph{Quantum cloning:} Let us now consider a few specific cases in some detail. We begin with the case of quantum cloning. 
Suppose that we have a two-dimensional quantum system (``qubit''), 
which in a certain instance is in the state \(|\varphi\rangle\) in the Hilbert space $ \mathbb{C}^{2} $. The \emph{protocol} in this case, is 
to have two exact copies of the state \(|\varphi\rangle\), and we suppose that we do not have any information about the input state other than that it is a qubit, so that we intend to consider a ``universal'' approximate cloning machine. So, we start with two qubits, one in the initial state $ |\varphi\rangle $, and the other in the fixed blank 
state $ |\varphi_{0}\rangle $.
The protocol is to have \(|\varphi\rangle \otimes |\varphi_{0}\rangle \longrightarrow |\varphi\rangle \otimes |\varphi\rangle\), which is quantum mechanically forbidden \cite{brisTi-poRe-Tapur-Tup}. 
The Bu{\v z}ek-Hillery (BH) quantum cloning machine \cite{jole-bhijte-moja-khub} is a quantum process that approximates this protocol. If $\{|0\rangle, |1\rangle\}$ is an orthonormal basis of the qubit, then the action of the BH machine can be expressed as
\begin{eqnarray}
|0\rangle |\varphi_{0}\rangle |M\rangle \rightarrow \sqrt{\dfrac{2}{3}} |00\rangle|\uparrow\rangle + \sqrt{\dfrac{1}{3}}|\Psi_{+}\rangle|\downarrow\rangle, \\
|1\rangle |\varphi_{0}\rangle |M\rangle \rightarrow \sqrt{\dfrac{2}{3}} |11\rangle|\downarrow\rangle + \sqrt{\dfrac{1}{3}}|\Psi_{+}\rangle|\uparrow\rangle,
\end{eqnarray}
where $ |M\rangle $ is the initial state of the cloning device, $ \left\{\left|\uparrow\right\rangle, \left|\downarrow\right\rangle\right\}$ forms an orthonormal basis of that device, and
\(|\Psi_{+}\rangle = \frac{1}{\sqrt{2}}(|01\rangle + |10\rangle)\)
is a two-qubit state in the joint Hilbert space $ \mathbb{C}^{2}\otimes \mathbb{C}^{2} $ of the original copy and its intended clone, spanned by $ \{|00\rangle,|01\rangle, |10\rangle, |11\rangle\} $.
 We use this machine to approximately clone an arbitrary qubit-state $ |\varphi\rangle = \alpha|0\rangle + \beta|1\rangle$, where $\alpha $, $ \beta $ are any two complex numbers with $ |\alpha|^{2} + |\beta|^{2} = 1$. Since ideal cloning is prohibited, the reduced state of the clone at the output is not the same as $ |\varphi\rangle $. Let us suppose that the clone is in the state $ \rho_{out} $. The machine is ``symmetric'' in that the output for the original copy is the same density matrix as that for the clone. Then one can use the ``local'' fidelity $ F_{\Phi_{BH}}=\langle \varphi|\rho_{out}|\varphi\rangle $ to measure the efficiency of the cloning machine. In case of the BH machine, the efficiency is well-known and is given by \(F_{\Phi_{BH}} = 5/6\). This value is independent of $ \alpha $ (or $ \beta $), i.e. the efficiency is $ 5/6 $ irrespective of the input state we want to clone; hence it is a universal symmetric ``isotropic'' approximate quantum cloning machine. Thus, averaging over input states  gives the same value of the average efficiency $ \overline{F}_{\Phi_{BH}} $.

Let us now inquire about the quantum process randomness of the BH machine from Eq. (\ref{eq:randomness}).
We find it to be zero irrespective of the input state, so that the average quantum process randomness $ \overline{Q}_{\Phi_{BH}} $ also vanishes.

Towards maximizing the efficiency and minimizing the randomness, 
we can consider the minimized value of the ratio $ \overline{Q}_{\Phi}/\overline{F}_{\Phi} $ as a figure of merit to decide the performance of a quantum process $ \Phi $. However, this cannot be done when either of $ \overline{Q}_{\Phi} $ or $ \overline{F}_{\Phi} $ becomes zero. In that case, instead of $ \overline{Q}_{\Phi}/\overline{F}_{\Phi} $, we can minimize $(\overline{Q}_{\Phi} - \overline{F}_{\Phi}) $ to optimize the performance. In case of the BH machine, it is the latter option that must be used and its value is $ -\frac{5}{6} $.


Let us now move on from the ``universal'' cloning of a qubit, i.e., approximate quantum cloning of an arbitrary qubit state, to the case of state-dependent cloning. In state-dependent cloning, the cloning process is optimized for a given ensemble of states. In particular, we consider an ensemble consisting of two states of a qubit which are, in general, non-orthogonal to each other. For such an ensemble, an ``optimal'' cloning machine, that maximizes the average quantum process efficiency, was constructed in \cite{bruss} by Bru{\ss} \emph{et al.} We now try to find whether this machine remains optimal when we minimize the ratio of average randomness to average efficiency. So, let us suppose that \(|a\rangle\) and \(|b\rangle\) are two pure non-orthogonal states of a qubit, given by
\begin{equation}
|a\rangle = \cos\theta|0\rangle + \sin\theta|1\rangle,\\
|b\rangle = \sin\theta|0\rangle + \cos\theta|1\rangle
\end{equation}
where the ``angle'' between $ |a\rangle $ and $ |b\rangle $ is $ \cos^{-1}(\sin2\theta) $, $ 0 \leq \theta \leq \pi/4 $. The state of the qubit which we wish to clone is either $ |a\rangle $ or $ |b\rangle $, occuring with equal probabilities. The states of the joint system consisting of the original copy and the clone, obtained after cloning of $ |a\rangle $ and $ |b\rangle $, are respectively $ |\alpha\rangle $ and $  |\beta\rangle $, given by
\begin{eqnarray}
|\alpha\rangle = U|a\rangle|0\rangle, \\
|\beta\rangle = U|b\rangle|0\rangle,
\end{eqnarray} where $ U $ is the unitary operator performing the cloning operation and $ |0\rangle $ is the initial state of the blank qubit (clone), which is then transformed into an approximate clone of the original qubit. The efficiency of this machine is manifested in the fidelity of the cloned states $ |\alpha\rangle $ and $ |\beta\rangle $ with respect to the states $ |a\rangle|0\rangle $ and $ |b\rangle|0\rangle $ respectively. This quantity is the average ``global'' fidelity of the cloning machine, and is given by \cite{bruss}
\begin{equation}
\label{global-fidelity}
\overline{F}_{\Phi_{C}}=\frac{1}{2}(|\langle \alpha|aa\rangle|^{2} + |\langle \beta|bb\rangle|^{2}).
\end{equation}

Let us now derive an analytical expression for the average quantum process randomness associated with such a state-dependent cloning machine. Eq. (\ref{dhush-sssala}) in this case reads as 
\begin{eqnarray}
\overline{Q}_{\Phi_{C}}=\dfrac{1}{2}\big(|\langle a|\langle a|\alpha\rangle| \sqrt{1-|\langle a|\langle a|\alpha\rangle|^{2}} +  \nonumber \\
|\langle b|\langle b|\beta\rangle| \sqrt{1-|\langle b|\langle b|\beta\rangle|^{2}}\big).
\end{eqnarray}


If $ \phi $, $ \gamma $, and $ \delta $ are the angles between $ |a\rangle|a\rangle $ and $ |b\rangle|b\rangle $, $ |\alpha\rangle $ and $|\beta\rangle$, and $ |a\rangle|a\rangle $ and $ |\alpha\rangle $ respectively (Fig. 1 (left panel)), then $ \phi = \cos^{-1}(\sin^{2} 2\theta) $ and $ \gamma = \cos^{-1}(\sin 2\theta) $(see \cite{bruss}), by using the unitarity of $ U $, assuming that all the vectors before and after cloning, remain in one plane. The angle between $ |\beta\rangle $ and $ |bb\rangle $ is $ \cos^{-1}(\phi-\delta-\gamma )$. Then we have
\begin{eqnarray}
\label{fidelity}
\overline{F}_{\Phi_{C}}=\frac{1}{2}\big(|\cos\delta|^{2}+|\cos(\phi-\delta-\gamma)|^{2}\big),
\end{eqnarray}
while
\begin{eqnarray}
\label{randomness1}
\overline{Q}_{\Phi_{C}}&=&\dfrac{1}{2}\big( |\cos\delta|\sqrt{1-\cos^{2}\delta}  \nonumber \\
&& + |\cos(\phi - \delta - \gamma)|\sqrt{1-\cos^{2}(\phi - \delta - \gamma)}\big) \nonumber \\
&=& \frac{1}{4}( |\sin 2\delta| + |\sin (2(\phi -\delta -\gamma))| ).
\end{eqnarray}

It was shown in \cite{bruss} that the average efficiency in this scenario is maximized when $ \delta=\frac{1}{2}(\phi - \gamma) $, i.e. when the angle between $ |aa\rangle $ and $ |\alpha\rangle $ is equal to that between $ |bb\rangle  $ and $ |\beta\rangle $. 

It is natural to ask what the average quantum process randomness will be when maximum average efficiency is reached. More importantly, we can ask whether the optimal machine that minimizes $ \overline{Q}_{\phi_{C}}/\overline{F}_{\Phi_{C}} $, is different from the one that maximizes $ \overline{F}_{\Phi_{C}}$. We begin by finding the machines that minimize $\overline{Q}_{\phi_{C}}$ (Eq. (\ref{randomness1})) and check whether it coincides with the ones maximizing of $ \overline{F}_{\Phi_{C}} $ (Eq. (\ref{fidelity})).

\textbf{Case \Rmnum 1}: $ \sin 2\delta > 0 $ and $ \sin 2(\phi-\delta-\gamma) > 0 $. Setting $ \dfrac{d\overline{Q}_{\Phi_{C}}}{d\delta} = 0$, we obtain
$ \frac{1}{2}(\cos 2\delta - \cos 2(\phi - \delta - \gamma)) = 0$, which in turn implies 
%
%
$ \delta = n\pi \pm (\phi - \delta - \gamma) $, 
$ n = 0, \pm1, \pm2, \ldots$.
Only $ n=0 $ is relevant in our case as other values give the same amount of randomness, so that $ \delta = \pm(\phi - \delta - \gamma)$. For the ``$+$'' sign, we have,
\begin{equation}
\label{valid}
\delta = \frac{1}{2}(\phi - \gamma),
\end{equation}
while for the ``$ - $'' sign,
\begin{equation}
\phi=\gamma.
\end{equation}
The second solution implies that the initial states $ |a\rangle $ and $|b\rangle$ are either equal (up to a phase) or orthogonal, so that exact cloning is possible. So, in case $ \phi \neq 0 $ and $ \phi \neq \pi/4 $, we are left with Eq.~(\ref{valid}). Now, 
$ \frac{d^{2}\overline{Q}_{\Phi_{C}}}{d\delta^{2}} = -\sin 2\delta - \sin 2(\phi - \delta - \gamma) < 0$.
So, at $ \delta = \frac{1}{2}(\phi - \gamma) $, $ \overline{Q}_{\Phi_{C}} $ reaches a local maximum.

\textbf{Case \Rmnum 2}: $ \sin 2\delta > 0 $ and $ \sin 2(\phi-\delta-\gamma) < 0 $.
Again, $ \dfrac{d\overline{Q}_{\Phi_{C}}}{d\delta} = 0$ implies that
$\frac{1}{2}(\cos 2\delta + \cos 2(\phi - \delta - \gamma)) = 0$, which in turn gives
%
%
$2\delta = (2n+1)\pi \pm 2(\phi - \delta - \gamma) $,
$ n = 0, \pm1, \pm2, \ldots$.
Again considering the  case $ n=0 $,
\begin{equation}
2\delta = \pi \pm 2(\phi - \delta - \gamma).
\end{equation}
For the ``$ + $'', we have
\begin{equation}
\delta = \pi/4 + \frac{1}{2}(\phi -\gamma),
\end{equation}
while for the ``$ - $'',
\begin{equation}
\label{not-valid}
\phi - \gamma = \pi/2.
\end{equation}
But $ \phi $ and $ \gamma $ are positive numbers (in radians) and can never have values higher than $ \pi/2 $, and so Eq.~(\ref{not-valid}) is not valid. Hence we have $ \delta = \pi/4 + \frac{1}{2}(\phi -\gamma) $.
For $ \sin 2\delta > 0 $, we need $ 0 < \frac{\pi}{4} + \frac{1}{2}(\phi -\gamma) < \pi $. Now, 
$ \frac{d^{2}\overline{Q}_{\Phi_{C}}}{d\delta^{2}} = -\sin 2\delta + \sin(2(\phi - \delta - \gamma)) < 0$. 
So, $ \delta = \frac{\pi}{4} + \frac{1}{2}(\phi -\gamma)$ also produces a local maximum.

\begin{figure}
\label{3plots}
\includegraphics[scale=0.40]{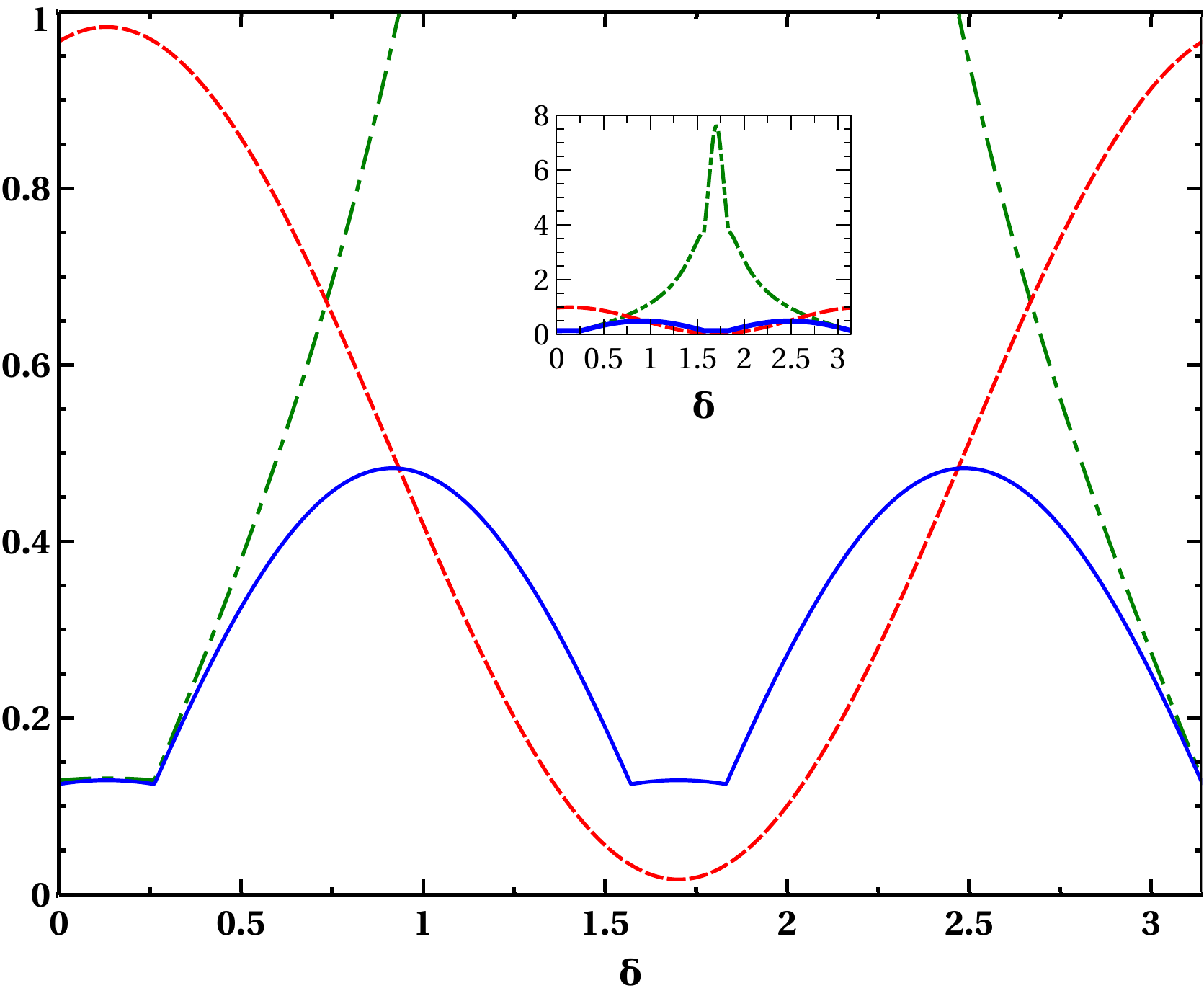}
\caption{Optimal state-dependent quantum cloning machines. The quantum fidelity 
$ \overline{F}_{\Phi_{C}} $ (red, dashed), quantum randomness $ \overline{Q}_{\phi_{C}} $ 
(blue, continuous) and their ratio, $\overline{Q}_{\phi_{C}}/\overline{F}_{\Phi_{C}} $ 
(green, dash-dotted), for state-dependent cloning, are plotted against $ \delta $, 
where we take $ \theta = \pi/8 $. The inset shows the behavior of the above three quantities 
with an extended  vertical axis. It turns out that 
the 
cloning machines which maximize the 
efficiency and which minimize the quantum randomness are different. Furthermore, the ones minimizing $\overline{Q}_{\Phi_{C}}/\overline{F}_{\Phi_{C}}$ coincide with those minimizing 
$\overline{Q}_{\Phi_{C}}$. All quantities plotted are dimensionless.}
\end{figure}

As we have studied the behavior of $ \overline{Q}_{\Phi_{C}} $ over the entire allowed region for $ \delta $, i.e. from $ 0 $ to $ \frac{\pi}{2} $ ($ \frac{\pi}{2} $ being the period of $ \overline{Q}_{\Phi_{C}} $) by the method of derivatives, and have been able to identify only the maxima of $ \overline{Q}_{\Phi_{C}} $, we can infer that the minima of $ \overline{Q}_{\Phi_{C}} $ lies at the boundary points of $ \delta $ and/or where $ \dfrac{d\overline{Q}_{\Phi_{C}}}{d\delta} $ is discontinuous. Further analysis reveals that the local minima, which are also the global ones, of $ \overline{Q}_{\Phi_{C}} $ are at $ \delta$=$0 $, $ \delta$=$\pi/2 $, and at an intermediate point $ \delta = \delta_0 \equiv \phi-\gamma $. We plot $ \overline{Q}_{\Phi_{C}} $ as a function of $ \delta $ for $ \theta=\pi/8 $ in Fig. 2. We find that $ \delta_{0}=\pi/12 $ in this case, and it is different from the value of $ \delta $ at which $ \overline{F}_{\Phi_{C}} $ is maximized. We comment here that when $ \delta=0 $, $ |\alpha\rangle $ is exactly $ |aa\rangle $ though $ |b\rangle $ may not be an exact clone of \(|\beta\rangle\). 
Focusing back to the $ \delta = \delta_{0} $ point, we therefore see that the best cloning machine is different, depending on whether we maximize the average efficiency or minimize the average quantum process randomness. The points where $ \overline{Q}_{\Phi_{C}}/\overline{F}_{\Phi_{C}} $ is minimized are the same where $ \overline{Q}_{\Phi_{C}} $ is minimized.


Now we assume that the vectors have a relative positioning as in Fig. 1 (right panel).
In this case also, the minima of $ \overline{Q}_{\Phi_{C}}/\overline{F}_{\Phi_{C}} $ coincide with the minima of $ \overline{Q}_{\Phi_{C}} $. However, the minimum value of $ \overline{Q}_{\Phi}/\overline{F}_{\Phi} $ in this case is higher than the minimum value obtained in case of Fig. 1 (left panel). 
The maximum fidelity obtained in this case also has a lower value than that obtained in the earlier case. So, we conclude that the earlier picture provides us with the optimal quantum cloning machines, irrespective of whether considered with respect to efficiency, or quantum randomness, or their ratio.

Lastly, we note that if in the cloning protocol, we have the information that the input is chosen from a set of mutually orthonormal vectors, exact cloning will become quantum mechanically 
possible, so that the average efficiency becomes unity, while the average randomness  vanishes. 


\emph{Quantum teleportation:} Let us now move on  to another specific case, viz. that of quantum teleportation. Suppose that the protocol is to teleport the quantum state \(|\varphi\rangle\) from a sender, called Alice, to a receiver, Bob, and 
that we have no information about the state to be sent, except that it is a qubit. 
To begin, suppose that the resources available are a single copy of a Bell state, say 
 $\frac{1}{\sqrt{2}}(|00\rangle + |11\rangle)$, 
shared between the sender and the receiver, and a finite amount of classical communication between them.
The celebrated quantum teleportation process \cite{bennett} achieves this protocol exactly, whereby the efficiency in this case is unity and the randomness is vanishing. The process consists in a measurement in the Bell basis by Alice on the joint Hilbert space formed by her part of the shared state and the state 
to be teleported, classical communication of the measurement result from Alice to Bob, and a unitary rotation at Bob's end that depends on the received classical 
information. The Bell basis is formed of the four orthonormal states \(|\Phi_{\pm}\rangle = \frac{1}{\sqrt{2}}(|00\rangle \pm |11\rangle)\), 
\(|\Psi_{\pm}\rangle = \frac{1}{\sqrt{2}}(|01\rangle \pm |10\rangle)\).


%
%

Next, let us suppose that among the resources available to perform the teleportation protocol, we replace the Bell state 
with the Werner state \cite{Braunschweig}, $\rho_W = p |\Phi_+\rangle \langle \Phi_+| + \frac{1-p}{4} I_4$,
between Alice 
and Bob, where \(I_4\) is the identity operator on the two-qubit Hilbert space, and where $ p \in [-1/3, 1] $. Following the same quantum process as when the shared state is \(|\Phi_+\rangle\), 
the teleported state $\rho_{\varphi}$, 
obtained at Bob's end, is given by
\begin{align}
\label{raat-koto-holo}
\rho_{\varphi} 
= p |\varphi\rangle \langle \varphi | + \frac{1-p}{2} I_2,
\end{align}
where 
\(I_2\) is the identity operator on the qubit Hilbert space.
 Though $\rho_{\varphi}$ is different 
for different $|\varphi\rangle$, $\langle \varphi | \rho_{\varphi} | \varphi\rangle = \frac{1+p}{2}$ depends only on the noise parameter $p$ in the Werner state. 
Also, $\langle \varphi | \rho^2_{\varphi} | \varphi\rangle = p+\frac14 (1-p)^2$ is independent of $|\varphi\rangle$. 
Hence, for the teleportation process with a shared Werner state, the quantum process randomness vanishes.

Let us now consider the teleportation protocol where the shared state -- between Alice and Bob -- is
 the noisy nonmaximally entangled state 
\(\rho_W^G=p|\eta\rangle \langle \eta| + \frac{1-p}{4}I_{4}\), where
\(|\eta\rangle=\cos\theta |00\rangle + \sin\theta |11\rangle\), with \(\theta \in [0,\pi/4]\). In this case, pursuing the teleportation process,
we find that for an input $ |\varphi\rangle = \alpha|0\rangle + \beta|1\rangle \in \mathbb{C}^{2}$, where $ \alpha = \cos(\frac{\mu}{2})e^{i\nu/2} $ and $ \beta = \sin(\frac{\mu}{2})e^{-i\nu/2} $, with $ \mu \in [0,\pi] $ and $ \nu \in [0,2\pi) $, the output, in the $\{|0\rangle, |1\rangle\}$  basis, is
\begin{center}
$ \rho_{\varphi} =$\(
p \begin{bmatrix}
   |\alpha|^{2} & \alpha\beta^{*}\sin 2\theta \\
    \alpha^{*}\beta\sin 2\theta & |\beta|^{2}
  \end{bmatrix} + \frac{1-p}{2}I_2
  \).
\end{center} 
Note that we are using the same notation for the output as in Eq. (\ref{raat-koto-holo}). The quantum process efficiency is given by $ F_{_{\Phi_{T}}}=\frac{1+p}{2}-2p(1-\sin 2\theta)|\alpha\beta|^{2} $, and the average quantum process efficiency is $ \overline{F}_{\Phi_{T}}=\frac{1+p}{2} - \frac{p}{3}(1-\sin 2\theta) $. The corresponding quantum process randomness is 
$ Q_{\Phi_{T}}=\frac{p}{4}(1-\sin 2\theta)|\sin 2\mu| $, so that $ \overline{Q}_{\Phi_{T}} = \frac{p}{6}(1-\sin 2\theta)$. Thus while there is finite nonzero randomness when the shared state between Alice and Bob is unentangled ($ \theta=0 $ case), the randomness vanishes when the shared state is maximally entangled.

Along with the fidelity for the quantum teleportation process (which we have referred here as the average quantum process efficiency), Ref. \cite{fidelity-dev} considers  
an interesting quantity that quantifies the variation of the fidelity as the input 
\(|\varphi\rangle\) moves over the relevant set \(\mathcal{S}\). The quantum process randomness is at a more ``atomic'' level, 
in the sense that it can be
    present even for a given \(|\varphi\rangle\). Once this quantum process randomness has
    been considered (for a given \(|\varphi\rangle\)), one is still left with the variation in that randomness as we move 
    over different \(|\varphi\rangle\) in \(\mathcal{S}\). This
    variation can now be quantified by the average (using e.g. the mean) and
    the spread (using e.g. the standard deviation).

To summarize, we have argued that the fidelity of a quantum process should be considered along with an associated quantum process randomness. We provided a measure for the proposed characteristic for an arbitrary quantum process. We analyzed the properties of the measure in approximate cloning and teleportation processes. In case of cloning, we found that the approximate quantum state-dependent machine that minimizes the process randomness or minimizes the ratio of process randomness and fidelity is different from the one that maximizes the fidelity.

\emph{Acknowledgment:} The research of SD was supported in part by the INFOSYS
scholarship for senior students.

\end{document}